\documentclass[twocolumn]{aastex62}
\usepackage{epstopdf}
\usepackage{wrapfig}
\usepackage{lipsum} 
\usepackage[utf8]{inputenc}
\usepackage{amsmath}
\usepackage{natbib}
\bibpunct{(}{)}{;}{a}{}{,}

\usepackage{color}
\definecolor{darkgreen}{RGB}{0,142,128}
\definecolor{darkblue}{RGB}{0,100,170}
\usepackage{hyperref}
\hypersetup{colorlinks,citecolor=blue,linkcolor=blue}

\graphicspath{{./}{figures/}}

\shorttitle{Tearing mode and PDS in the slow wind}
\shortauthors{Réville et al.}
\begin{document}

\title{Tearing instability and periodic density perturbations in the slow solar wind}

\author[0000-0002-2916-3837]{Victor Réville}
\affiliation{IRAP, Universit\'e Toulouse III - Paul Sabatier,
CNRS, CNES, Toulouse, France}

\author[0000-0002-2381-3106]{Marco Velli}
\affil{UCLA Earth Planetary and Space Sciences Department, LA, CA, USA}

\author[0000-0003-4039-5767]{Alexis P. Rouillard}
\affiliation{IRAP, Universit\'e Toulouse III - Paul Sabatier,
CNRS, CNES, Toulouse, France}

\author[0000-0001-6807-8494]{Benoit Lavraud}
\affiliation{IRAP, Universit\'e Toulouse III - Paul Sabatier,
CNRS, CNES, Toulouse, France}

\author[0000-0003-2880-6084]{Anna Tenerani}
\affil{University of Texas, Austin, TX, USA}

\author[0000-0002-2582-7085]{Chen Shi}
\affiliation{UCLA Earth Planetary and Space Sciences Department, LA, CA, USA}

\author[0000-0002-9630-6463]{Antoine Strugarek}
\affiliation{AIM, CEA, CNRS, Université Paris-Saclay, Université Paris Diderot, Sorbonne Paris Cité, F-91191 Gif-sur-Yvette, France}

\begin{abstract}
In contrast with the fast solar wind, that originates in coronal holes, the source of the slow solar wind is still debated. Often intermittent and enriched with low FIP elements -akin to what is observed in closed coronal loops- the slow wind could form in bursty events nearby helmet streamers. Slow winds also exhibit density perturbations which have been shown to be periodic and could be associated with flux ropes ejected from the tip of helmet streamers, as shown recently by the WISPR white light imager onboard Parker Solar Probe (PSP). In this work, we propose that the main mechanism controlling the release of flux ropes is a flow-modified tearing mode at the heliospheric current sheet (HCS). We use MHD simulations of the solar wind and corona to reproduce realistic configurations and outflows surrounding the HCS. We find that this process is able to explain long ($\sim 10-20$h) and short ($\sim 1-2$h) timescales of density structures observed in the slow solar wind. This study also sheds new light on the structure, topology and composition of the slow solar wind, and could be, in the near future, compared with white light and in situ PSP observations.
\end{abstract}


\section{Introduction} 
\label{intro}

In the recent years, the usual dichotomy between a fast ($> 600$ km/s) and a slow ($\sim 350-400$ km/s) wind has been challenged as the main discriminating factor to identify the source of the solar wind plasma. A more accurate picture could be described as follows: a rather steady component (often fast but not always), coming from open regions, or coronal holes, is associated with a well developed turbulent spectrum of Alfvénic fluctuations that actively contribute to the acceleration of the solar wind through an extended wave pressure and dissipation at kinetic scales \citep{Belcher1971,Leer1982,Velli1991,VerdiniVelli2007,Verdini2009}. As a consequence of the extended energy deposition, the wind will reach higher speeds than a purely thermally driven outflow. Yet, the terminal velocity is fundamentally dependent on the processes occurring in the low corona, and a dense, slow wind can still emerge from a coronal hole if a strong expansion and/or heating is present below the sonic point \citep[see, e.g.,][]{Velli2010}. 

A second component, more intermittent and dynamic, originates from or nearby streamers, i.e. closed coronal structures. This characterization of the "slow" wind is backed by composition analyses, which show a strong enrichment in low First Ionization Potential (FIP) elements \citep[see, e.g.,][]{Laming2004}. Some dynamical processes must then allow for exchange between the closed structures and the outflow \citep{Antiochos2012,HigginsonLynch2018}. Moreover, the wind surrounding the heliospheric current sheet (HCS) reveals density perturbations that are not observed in the fast wind. These density perturbations have been shown to be periodic, with high power concentrated around a 90 minute period, coherently over many different instruments at various places in the inner heliosphere \citep{Viall2010,Viall2015}. \citet{SanchezDiaz2017b,SanchezDiaz2019} have further identified longer periods of 10 to 20 hours driven by the release of extended structures along with the aforementioned substructures of ten times smaller periods. Recent solar wind measurements by the \textit{Parker Solar Probe}, combined with remote-sensing observations taken near 1 au, provide additional support to the idea that the slow wind comes in at least two states with different bulk properties and levels of variability \citep{Rouillard2020ApJSa}.

The suggestion that reconnecting plasmoids at the tips of helmet streamers were a slow wind source had already been made by \citet{Sheeley1997, Wang1998}. The study of \citet{SanchezDiaz2017a} has recently shown using observations of the SOHO and STEREO spacecraft, that the formation of density structures in the corona were associated with inflows that support formation by magnetic reconnection. \citet{Endeve2003,Endeve2004}, while studying simple dipolar configurations, unveiled an instability that leads to magnetic reconnection and the periodic release of flux ropes from the tips of the streamers. These works identified thermal processes as the origin of the streamer's instability, the thermal conduction coefficient being the critical parameter to make the instability vanish. The periods obtained in these works are of the order of 15 hours, and thus can only account for the longer timescales observed in the inner heliosphere, which suggests that another mechanism might be at play.

The HCS is pinched at the tips of streamers, which can lead to a tearing instability \citep{Furth1963,Biskamp1986,VelliHood1989,Loureiro2007,PucciVelli2014, Tenerani2015}. \citet{Einaudi1999} modeled the region above the cusp of a helmet streamer as a current sheet embedded in a broader wake flow. The combined tearing Kelvin-Helmholtz instability was shown to induce the acceleration of density-enhanced magnetic islands. The model was further developed by \citet{Rappazzo2005}, who showed how the diamagnetic plasmoid expulsion in a spherical geometry would lead to a rapid plasmoid acceleration profile. However, these works started with a finite thickness current sheet and did not take into account either the specific geometry of the helmet streamer cusp, nor the natural thinning of the sheet arising from the converging plasma flow, although they did comment on the role such convergence might play in the evolution. 

In this letter, we revisit the instability of streamers using a resistive MHD model of the solar corona and wind. We investigate the effect of the Lundquist number $S$ on the streamers' stability and the periodicity of the structures released in the slow solar wind. We show that, for sufficiently high $S$, a tearing instability is launched in the HCS and that its scaling properties with $S$ are consistent with the ideal tearing scenario \citep{PucciVelli2014,Tenerani2015}. We observe two main periodicities in the simulations: a small one related to the fastest growing tearing mode, and a longer one associated with cycles of linear onsets of the instability. Using the ideal tearing scalings to very high $S$ typical of the solar corona, we show that these periodicities are fully compatible with the timescales of density structures observed in the slow solar wind.

\section{MHD model and solar wind parameters}
\label{sec:numerics}

We use the Alfvén wave turbulence driven MHD model presented in \citet[][]{Reville2020ApJS}, itself based on the PLUTO code \citep{Mignone2007}, and refer the reader to this paper for all the details on the numerical schemes and equations solved. For this study, we perform 2.5D simulations in spherical geometry of a dipolar configuration, corresponding to a solar minimum case. We focus on the properties of the HCS, located around $\theta = \pi/2$. For the need of this particular study, some improvements have been brought to the numerics of the code. First, we use a HLLD Riemann solver \citep{MiyoshiKusano2005}, which handles discontinuities and sharp gradients better than the previously used HLL solver \citep{Einfeldt1988}. In particular, our studies of reconnecting current sheets using both solvers have shown that only the HLLD was able to correctly capture the growth of a tearing instability. Secondly, we use a constrained transport method \citep{BalsaraSpicer1999} to maintain $\nabla \cdot \mathbf{B} = 0$ at machine accuracy. Finally, explicit resistivity is introduced to ensure a good control of the Lundquist number close to the current sheet. The simulations are integrated on a non uniform grid, strongly refined around the current sheet with $\Delta \theta = 10^{-3}$ for $\theta \in [\pi/2-0.1,\pi/2+0.1]$. The radial grid is stretched, although we maintain an almost constant $\Delta r = 10^{-2}$ up to $20 R_{\odot}$, while the solution is extended up to $50 R_{\odot}$ with a coarser grid. The aspect ratio of the cells $r\Delta \theta/\Delta r$ is close to unity between $1$ and $15 R_{\odot}$, which encompasses the growth region of the tearing instability that we will characterize in the following section. Although the resolution might seem coarse for this kind of problem, we use a 4th order spatial scheme obtained by a parabolic reconstruction, which effectively ensure results that are consistent with linear theory up to $S \sim 10^6$ (see section \ref{sec:tearing}).

\begin{figure}
\center
\includegraphics[width=3.in]{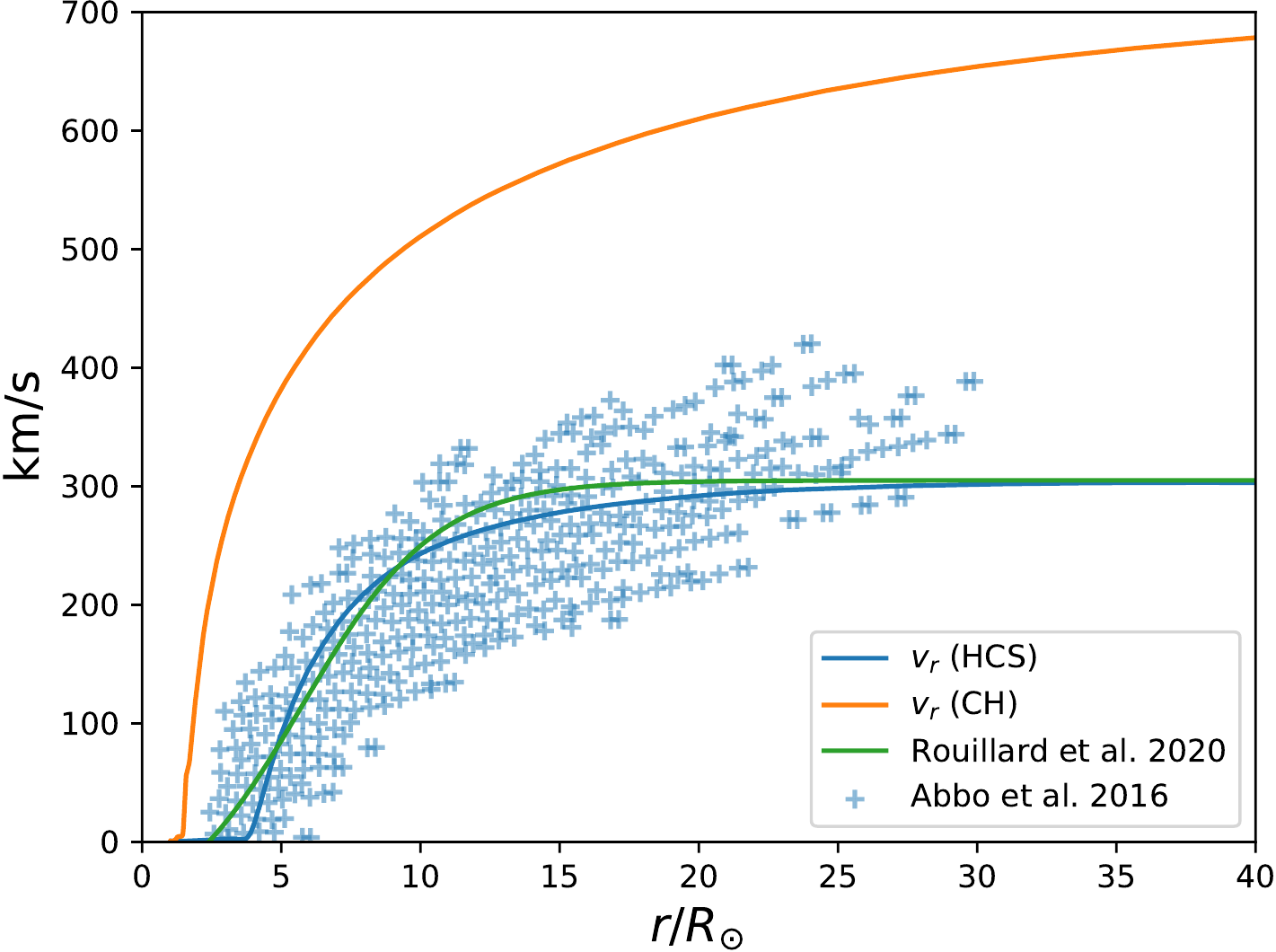} \\
\caption{Wind speed obtained in the simulation in a coronal hole and at the HCS. The slow wind (in blue) obtained near the current sheet is compared with the speed of observed density structures from SOHO/LASCO and STEREO/COR \citep{Abbo2016}, as well as a numerical fit \citep{Rouillard2020ApJSb}.}
\label{fig:Abbo}
\end{figure}

The model has already shown a very good agreement with bulk solar wind properties at the \textit{Parker Solar Probe} first perihelion \citep[see][]{Reville2020ApJS}. There are essentially $4$ input parameters in the model: the amplitude of Alfvén waves at the coronal base $\delta v_{\odot} = 30$ km/s, the base density $\rho_{\odot} = 5 \times 10^{-16}$ g.cm$^{-3}$, the correlation length of the turbulence at the base of the corona $\lambda_{\odot} = 0.025 R_{\odot}$, and the input magnetic field. These input parameters are very close to the ones used in the study of the first PSP perihelion, except for the input magnetic field which was then given by a solar magnetogram. Instead, we use here a purely dipolar field of $5$ Gauss at the equator which reproduces typical fast and slow wind bulk properties at solar minimum.

In Figure \ref{fig:Abbo}, we compare the wind speed profile obtained in our simulations with the speed of density structures observed in the inner heliosphere with the white-light instruments SOHO/LASCO and STEREO/COR \citep{Sheeley1997,Abbo2016}. The fast wind speed in orange is obtained at high latitudes, while the slow wind speed profile is extracted at the HCS in our model. The slow wind speed is very close to the fit made to the observations in \citet{Rouillard2020ApJSb} and fully consistent with the speeds of the slow wind. The whole system of flows and density fields around the HCS is obtained self-consistently with the Alfvén wave driven heating model and ensures that realistic physical conditions are met at HCS, where a tearing instability is triggered, as described in the following sections.

\section{Onset of a tearing instability in the HCS}
\label{sec:tearing}

\begin{figure}
\center
\includegraphics[width=3in]{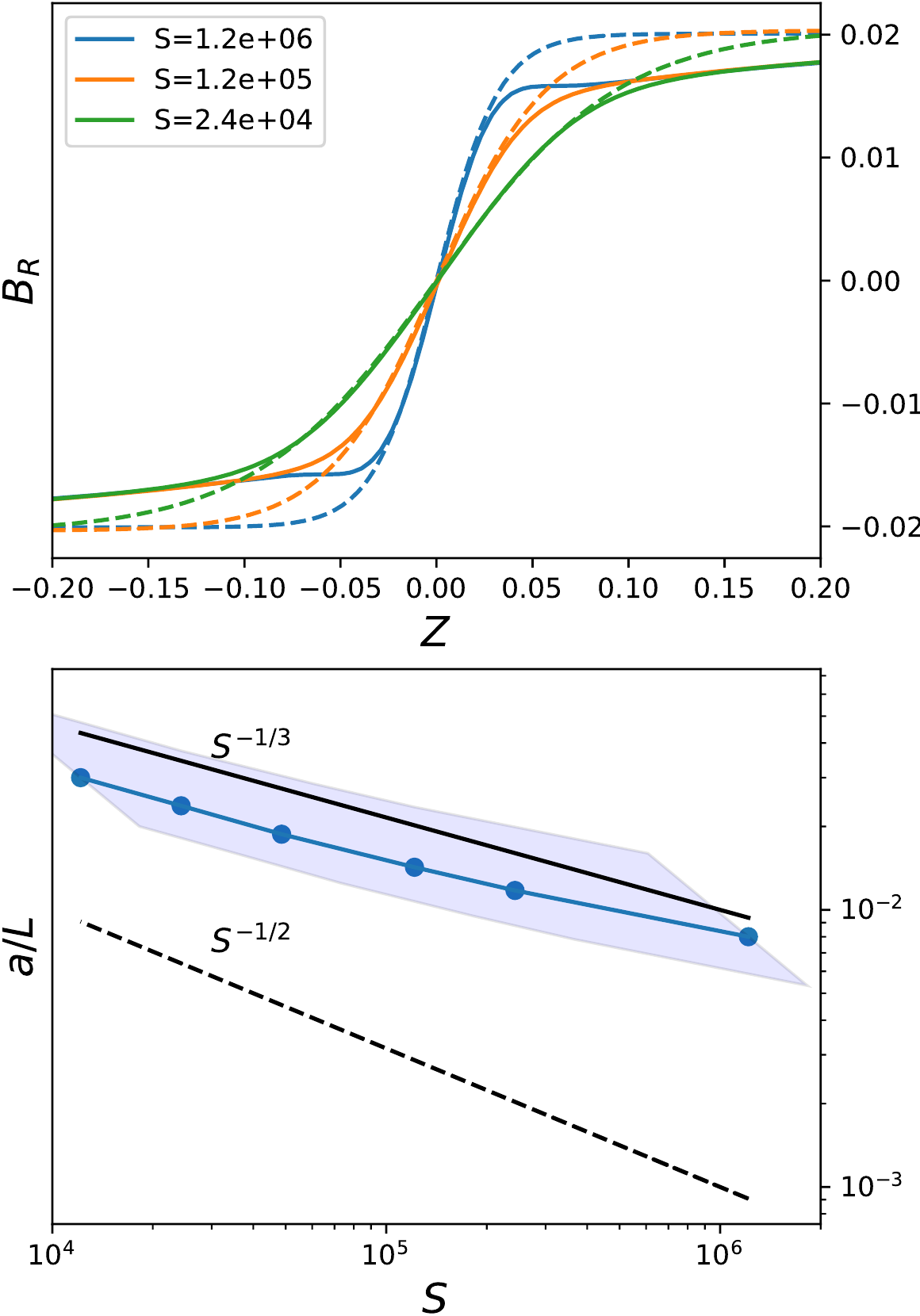} \\
\caption{Top panel: cuts of the radial field structure close to the HCS for three of the simulations. Dashed lines shows formulation \eqref{eq:harris} using $|A| = |B_Z (R=5,Z = 2)|$ and the best fit for $a$. Bottom panel: aspect ratio $a/L$ of the current sheet as a function of the Lundquist number for all six simulations. The plain blue line is obtained using $L=4 R_{\odot}$ and the shaded area spans $L \in [2 R_{\odot}, 6 R_{\odot}]$. The critical aspect ratio $a/L = S^{-1/3}$ is shown in plain black while the Sweet-Parker aspect ratio $a/L = S^{-1/2}$ is represented with the dashed line.}
\label{fig:thickness}
\end{figure}

This letter presents six simulations in which the Lundquist number $S=L V_A/\eta$ varies around the HCS. The value of $L = 4 R_{\odot}$, the characteristic length, and $v_A = v_{\mathrm{kep}} \sim 437$ km/s are the same in all simulations and correspond to the growth region of the tearing instability and the Alfvén speed away from the current sheet. We only change $\eta$ so that $S$ varies between $10^4$ and $10^6$. At the beginning of the simulations, the magnetic field is a pure dipole and there is no current sheet. The first phase hence consists in the creation and the thinning of the HCS around $ \theta = \pi /2$ within a solar wind flow. In addition to the flow, latitudinal, equatorward Lorentz forces act to thin further the HCS \citep[see, e.g.][]{Reville2017}. The thinning process takes roughly $90$ hours in all our simulations.

Studies of current sheet reconnection usually starts with a given equilibrium of magnetic field and velocities. The Harris current sheet, for example, sets the field along the sheet, $B_R$ in a cylindrical coordinate system $(R,Z,\Phi)$, to be:

\begin{equation}
    B_R (Z) = A \tanh \left( \frac{Z}{a} \right),
    \label{eq:harris}
\end{equation}
where $a$ is the thickness of the current sheet. Recent works \citep{PucciVelli2014,Tenerani2015} have tried to assess what aspect ratios $a/L$ could be physically constructed and lead to fast reconnection. In particular, \citet{PucciVelli2014} showed that Sweet-Parker current sheets would generally be impossible to create because of the diverging (infinite) growth rates for the tearing instability as $S \rightarrow +\infty$. They concluded that current sheets would disrupt once a limiting inverse aspect ratio  
scaling as $a/L \propto S^{-1/3}$ would be reached, at which point the tearing mode growth rate becomes independent of $S$ and of the same order as that of ideal MHD dynamics.

In the top panel of Figure \ref{fig:thickness}, we show the structure of the current sheet obtained in the simulations at the end of the thinning phase. The thickness of the current sheet is computed using equation \eqref{eq:harris} and the value $|A| = |B_Z (R=5, Z=2)|$, the amplitude of the field away from the current sheet. As the Lundquist number increases the current sheet thins and we give the obtained scaling in the bottom panel of Figure \ref{fig:thickness}. After the thinning process, different evolutions are observed depending on the value of $S$. For the lowest $S = 1.2 \times 10^4$, the current sheet is stable. For $S= 2.4 \times 10^4$, a regime of steady reconnection sets in and for all higher $S$, a tearing instability is triggered. The bottom panel shows the aspect ratio $a/L$ and compares with theoretical scalings. The plain blue line is obtained with $L=4 R_{\odot}$ while the shaded area represents the interval $L \in [2 R_{\odot},6 R_{\odot}]$. The aspect ratios $a/L$ that we get are interestingly always smaller than $S^{-1/3}$. The reason for this smaller ratio might reasonably be found in the presence of a small normal component of the magnetic field, whose stabilizing effect can lead to smaller inverse aspect ratios for fast ideal growth rates shown by \citet{PucciVelli2019}. Nonetheless, at large $S$, the aspect ratio $a/L$ approaches the $S^{-1/3}$ scaling, diverging clearly from the well-known Sweet-Parker scaling $S^{-1/2}$.

\begin{figure}
\center
\includegraphics[width=3.5in]{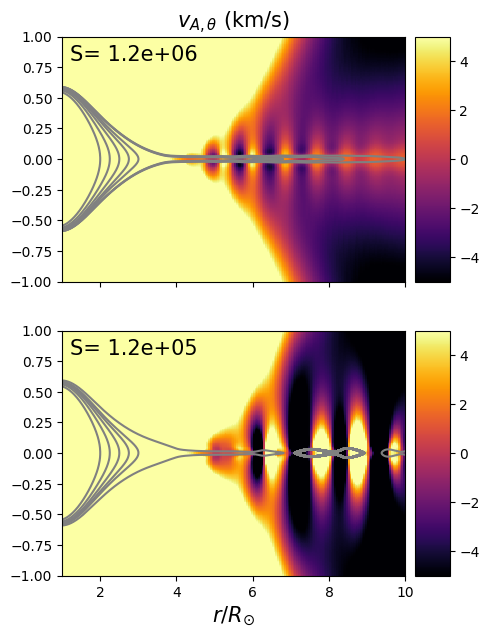} \\
\caption{Snapshot of the fastest growing mode of a tearing instability in the current sheet. The background color is the latitudinal Alfvén speed $v_{A,\theta}$. The Lundquist number $S$ is equal to $1.2 \times 10^6, 1.2 \times 10^5$ in the top and the bottom panel respectively. The aspect ratio is not respected to ease the visualization of the structures at the HCS.}
\label{fig:tearing}
\end{figure}

Figure \ref{fig:tearing} shows the growth of the tearing instability for two unstable cases. The quantity in the background color is the latitudinal Alfvén speed $v_{A,\theta}$, which is characteristic of the tearing mode. Using the Alfvén speed instead of the magnetic field component across the sheet is a way of accounting for the expansion of the wind. Analysis of the toroidal electric field yields similar results. The snapshots are taken at the end of the linear growth phase of the fastest growing mode $k_m$ in the simulations. As $S$ increases, we can see that $k_m$ increases (or that the characteristic wavelength decreases). The growth region is found for all simulations to be between $5$ and $9 R_{\odot}$, which explains \textit{a posteriori} our choice of $L = 4 R_{\odot}$.

\begin{figure}
\center
\includegraphics[width=3.0in]{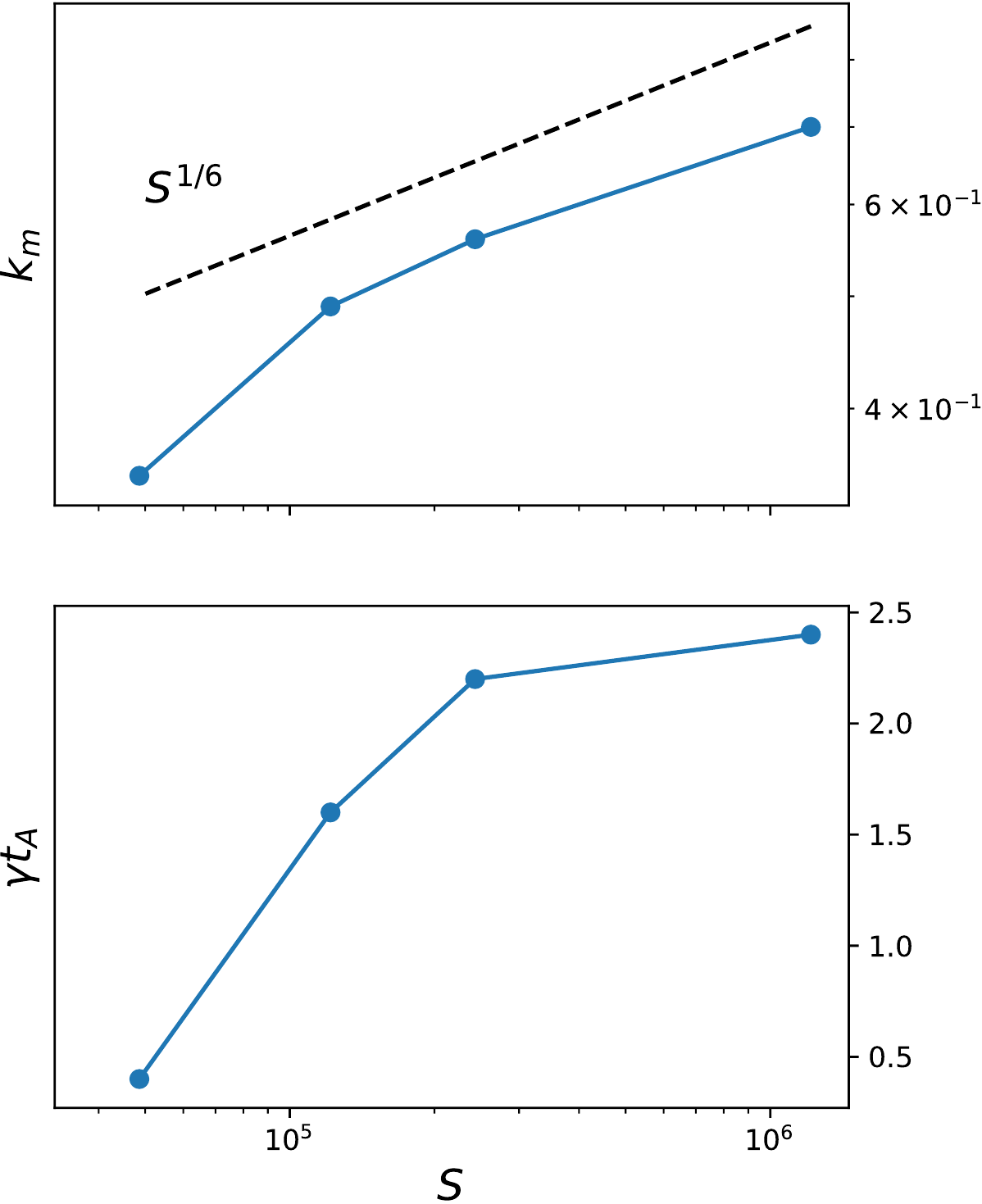} \\
\caption{Top panel: fastest growing mode $k_m$ as a function of the Lundquist number $S$. The asymptotic regime $k_m \propto S^{1/6}$ is shown by the black dashed line. The corresponding growth rate, normalized by $t_A = L/v_A$, is shown in the bottom panel.}
\label{fig:growth}
\end{figure}

According to the ideal tearing scenario \citep{PucciVelli2014,Tenerani2015}, when the current sheet aspect ratio $a/L \sim S^{-1/3}$, we get the following asymptotic scaling for the fastest growing mode:
\begin{equation}
    k_m L \propto S^{1/6}, \; \gamma t_A = C,
\end{equation}
with the constant $C\simeq 0.62$ for the Harris equilibrium profile and taking precisely $a/L=S^{-1/3}$. Figure \ref{fig:growth} shows the computed scaling of $k_m$ for the fastest growing mode and the associated rate for all unstable cases. As discussed earlier, at our values of $S$ the fastest growing mode has a growth rate that increases with $S$, and although our simulations only reach $S \sim 10^6$, the slope of $k_m (S)$ is seen to decrease towards the predicted limiting scaling, illustrated by the black dashed line. The growth rate shown here also tends to converge towards a fixed value $\gamma t_A \sim 2.5$, or at least does not explode with increasing $S$. This value, higher than expected for Harris equilibria, could be the result of gradients in velocity and density which are subsequent to the realistic wind conditions met around the current sheet in the simulations. 

\section{Reconnection and periodic density perturbations}
\label{sec:pds}

\begin{figure}
\center
\includegraphics[width=3.5in]{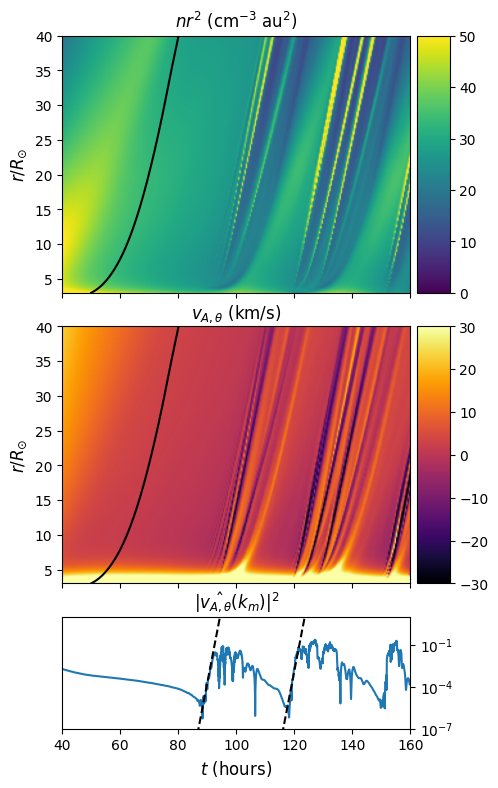} \\
\caption{Propagation of the flux ropes structures in the solar wind near the equator for the case $S=2.4 \times 10^5$.  The expansion compensated density ($n r^2$, top panel) and the transverse Alfvén speed ($v_{A, \theta}$, middle panel) are sliced near the HCS and stacked over time. Structures propagate at the slow wind speed, materialized by the characteristic curve in black. The growth of the fastest growing mode is shown in the bottom panel with an exponential fit at $\gamma t_A=2.2$, and exhibits several linear onsets corresponding to the release of structures in the corona.}
\label{fig:pbs}
\end{figure}

Magnetic islands, or flux ropes in three dimensions, created by the tearing instability described in the previous section, do create a local enhancement of the density and their propagation in the solar wind could lead to the in situ and white light observations of PDS. In Figure \ref{fig:pbs}, we show the propagation of the structures created by the tearing instability and the reconnection at the tip of the helmet streamer. The transverse Alfvén speed $v_{A,\theta}$ and the expansion compensated density $n r^2$ are taken at the HCS and stacked over time. The substructures of alternate signs in the Alfvén speed are the flux ropes created by the tearing mode. Density enhancements are associated with the edges of these flux ropes and reach about two to three times the local density, which is consistent with observations \citep[see, e.g.,][]{SanchezDiaz2017b}. The scales predicted by the tearing mode analysis of section \ref{sec:tearing} can be computed as follows: because the simulation with the largest Lundquist number is close to the asymptotic regime, we can scale up the fastest growing mode flux ropes at $S \sim 10^{12}$, which is the expected value in the solar corona:
\begin{equation}
    \lambda_m (S=10^{12}) = 1/k_m (S=10^6)\times (10^{-6})^{1/6} \sim 0.15 R_{\odot}. 
\end{equation}
This length scale does not account for the expansion and acceleration of the wind. Figure \ref{fig:pbs} shows the characteristic curve of the propagation of a passive scalar advected by the wind speed near the HCS. This curve matches closely the kinematics of the islands and periodic density perturbations shown in the colorplots. When released, likely by a combination of flow dragging and melon-seed forces \citep[see][]{Rappazzo2005}, density structures are advected by the slow solar wind. They gradually accelerate out to a heliocentric distance of 10-15 solar radii and then propagate at nearly constant speed. Moreover, we can see in Figure \ref{fig:pbs} that the structures grow as they propagate in the slow solar wind. This can be understood as a consequence of the solar wind acceleration in the growth region of the tearing mode. If these structures are pressure balanced and advected by the wind, two points initially separated by $d\lambda = (x_1 - x_0)$ will eventually be separated by 
\begin{equation}
    d \lambda' (t) = d\lambda + \int_t v(x_1(t)) -  v(x_0(t))dt,
    \label{eq:scaleexp}
\end{equation}
after a time $t$ \citep[see also][]{Bulanov1978,Shi2018}. We find using the slow wind speed profile used in Figure \ref{fig:Abbo} that between $3 R_{\odot}$ and $20 R_{\odot}$ a structure of characteristic scale $\lambda_m = 0.15 R_{\odot}$ can expand by a factor $50$ along the sheet. A straightforward integration of equation (\ref{eq:scaleexp}) for $x_0(0) = 4R_{\odot}$ and $d\lambda(0)=\lambda_m = 0.15 R_{\odot}$ gives $d\lambda = 2R_{\odot}$, out of the solar wind acceleration region. Provided that these structures propagate at some $300$ km/s, we get a typical period of $80$ minutes out of the solar wind acceleration region, which is very close to the periodicity of density structures reported in \citet{Viall2010,Viall2015}. Depending on the precise birth location of the flux rope, however, the growth will vary, as the expansion is an increasing function of the time spent in the acceleration region.

Finally, we also observe in Figure \ref{fig:pbs} a second periodicity that could be the longer one reported by the observations. The last panel of Figure \ref{fig:pbs} displays three episodes of linear onsets followed -for the first two- by a non-linear phase of the tearing mode, that are periodically evacuated by the solar wind outflow. During the non-linear phase, flux and plasma is transferred from the streamer to the HCS. The composition of the density structures thus should be close to the composition of the helmet streamer, which is consistent with observations. These episodes are separated by roughly $30$ hours, which is slightly more than observations at activity maximum, yet in the right order of magnitude \citep{SanchezDiaz2017b,SanchezDiaz2019}. This periodicity is of course controlled by the strength and duration of the non-linear stage. We can expect that as $S$ increases, and the size of the structures decreases, the solar wind flow will be able to evacuate more easily the perturbations, hence reducing the time interval between linear onsets. Further studies are however necessary to confirm this behaviour of the non-linear regime at high $S$.

\section{Discussions}
\label{sec:dis}

In this work, we have shown that a tearing instability was naturally occurring at finite resistivity in simulations of the HCS with realistic outflow velocity and density gradients. At low Lundquist numbers, the current sheet is stable and no reconnection is observed. Increasing $S$, the current sheet thins, and reconnection is triggered via a tearing mode. The current sheet before the onset of the instability is thinner than the ideal tearing scenario for low $S$ but the inverse aspect ratio approaches the scaling $a/L \propto S^{-1/3}$ as $S$ increases. The fastest growing mode and associated growth rate also reach an asymptotic regime which is consistent with the ideal tearing scenario, with $k_m \propto S^{1/6}$ and $\gamma t_A \sim 2.5$. Gradients and inhomogeneities in the simulations could explain the difference with the theoretical values predicted by \citet{PucciVelli2014}. Another reason might be the previously mentioned finite normal component in the neighborhood of the helmet streamer cusp. This has a stabilizing influence, requiring sheets to become thinner for tearing modes to occur on fast ideal timescales \citet{PucciVelli2019}.

The convergence towards an ideal regime allows some extrapolation at much larger $S$, characteristic of the conditions in the solar corona. We can expect that such a tearing mode for the actual HCS could give typical length-scales of about $0.15 R_{\odot}$ at the tip of the streamers. These flux ropes are then ejected and grow as pressure balanced structures inside the solar wind acceleration region. Accounting for this expansion, the length and timescales of flux ropes as well as the density structures that separate them are fully consistent with periods of a few hours observed in heliospheric images \citep{Viall2010,Viall2015} and in situ data \citep{Kepko2016,DiMatteo2019}.

This paper is mostly focused on the linear regime of the tearing instability. Our simulations show an interesting behaviour, where several linear growth phases of the tearing mode are evacuated with the solar wind. Such a recurring process could explain why the linear fastest growing mode has a peculiar status in the observations, as well as the longer periods of flux ropes events of 10 to 20h. The precise duration of these cycles is however controlled by the properties of the non-linear regime, and needs to be studied in further details. Our results suggest nonetheless that in the non-linear phase plasma initially trapped on coronal loops below helmet streamer and enriched in low FIP elements could be released in the slow solar wind.

Finally, it is important to stress here that our results differ quite fundamentally from the streamer's instability mechanism proposed by \citet{Endeve2003,Endeve2004}. Interestingly, while we do get siphon flows \citep[see, e.g.][]{CargillPriest1980} for resolutions equivalent to that of these latter works, we don't observe any thermal instabilities at the relatively high resolution used for our present dipolar study. Hence, although thermodynamic processes related to coronal heating are likely to be responsible for reconnection in the low corona, they may not be necessary to explain the observations of periodic density perturbations. Moreover, a tearing-induced reconnection is likely much less dependent on the scale and properties of the streamers, which is consistent with regular periodicities of solar wind density structures over the solar cycle \citep[see][]{Viall2009}. Analysis of early FIELDS and SWEAP data during the first PSP orbit seem to suggest that the whole heliospheric plasma sheet may be generated by the ensemble of reconnection exhausts originating from the tips of helmet streamers \citep{Lavraud2020ApJL}. In the near future the WISPR imager onboard PSP may be close enough to the streamer cusp region to provide global observations of the reconnection process.

\section{acknowledgement}
This research was funded by the ERC SLOW{\_}\,SOURCE project (SLOW{\_}\,SOURCE - DLV-819189) and the NASA Parker Solar Probe Observatory Scientist grant NNX15AF34G. VR thanks Fulvia Pucci for useful discussions. The authors are grateful to A. Mignone and the PLUTO development team. Simulations were performed on the Extreme Science and Engineering Discovery Environment \citep[XSEDE,][]{Xsede2014} SDSC based resource Comet with allocation number AST180027, and GENCI supercomputers (grant 20410133). XSEDE is supported by National Science Foundation grant number ACI-1548562. This study has made use of the NASA Astrophysics Data System.

\bibliographystyle{yahapj}
\bibliography{./biblio}

\end{document}